\documentclass[preprint2]{aastex}
\usepackage{color}
\usepackage{multirow}

\slugcomment{Accepted for publication in ApJ}

\shorttitle{X-Ray Stacking}

\begin{document}

\title{DETECTION OF IRON K$\alpha$ EMISSION FROM A COMPLETE SAMPLE
OF SUBMILLIMETER GALAXIES}

\author{Robert R. Lindner\altaffilmark{1}, 
        Andrew J. Baker\altaffilmark{1}, Alexandre Beelen\altaffilmark{2}, Frazer N. Owen\altaffilmark{3}, and Maria Polletta\altaffilmark{4}}

\altaffiltext{1}{Rutgers, The State University of New Jersey, 136 Frelinghuysen Road, Piscataway, 
NJ 08854-8019, USA; lindner@physics.rutgers.edu, ajbaker@physics.rutgers.edu}
\altaffiltext{2}{Institut d'Astrophysique Spatiale, Universit\'e Paris Sud 11 and CNRS, Orsay, France; alexandre.beelen@ias.u-psud.fr}
\altaffiltext{3}{National Radio Astronomy Observatory, P.O. Box O, Socorro, NM 87801, USA; fowen@nrao.edu}
\altaffiltext{4}{INAF Ð IASF Milano, via E. Bassini, 20133 Milan, Italy;  polletta@iasf-milano.inaf.it}

\begin{abstract}
We present an X-ray stacking analysis of a sample of
38 submillimeter galaxies with $\left< z \right>=2.6$ discovered 
at $\geq 4\sigma$ significance in the Lockman Hole North with the 
MAMBO array.  
We find a $5\sigma$ detection in the stacked soft band 
(0.5--$2.0\,\rm keV$) image, and no significant detection in the 
hard band (2.0--$8\,\rm keV$).  We also perform rest-frame 
spectral stacking based on spectroscopic and photometric redshifts
and find a $\sim 4\sigma$ detection of $\rm Fe\,K\alpha$ emission 
with an equivalent width of $\rm EW\gtrsim1\,\rm keV$.  The centroid of the 
$\rm Fe\,K\alpha$ emission lies near $6.7\,\rm keV$, indicating a 
possible contribution from highly ionized Fe\,XXV or Fe\,XXVI; there is 
also a slight indication that the line emission is  
more spatially extended than the X-ray continuum.  This is the 
first X-ray analysis of a complete, flux-limited sample of SMGs with 
statistically robust radio counterparts.

\end{abstract}


\section{Introduction}

Submillimeter galaxies (SMGs) are distant star-forming systems
with tremendous infrared luminosities 
($L_{\rm IR}$ [8--1000$\,\mu{\rm m}] \gtrsim 10^{12}\, L_{\odot}$).  In 
the (sub)millimeter waveband they are observable out to high 
redshifts due to the strong negative $K$-correction in the Rayleigh-Jeans 
regime of their thermal spectrum \citep[see, e.g.,  ][]{blai02}.  The 
prevalence of SMGs at $z> 1$ \citep{chap05} in combination with their
high rates of dust-obscured star-formation imply  that 
they may be responsible for the production of a significant 
fraction of all the stellar mass in present-day galaxies.  X-ray 
 \citep{alex03,alex05b} and mid-infrared 
 \citep{vali07, mene07, mene09, pope08} spectroscopy shows that 
SMGs frequently contain active galactic nuclei (AGN) as well as
powerful starbursts.  This connection
between star formation and accretion at high redshift may help
explain the black hole mass-bulge mass relation 
in present-day galaxies \citep[e.g., ][]{alex05a}.  However, 
it remains hard to determine the relative importance of accretion 
and star formation for the SMG population as a whole because 
of the challenge of assembling large, statistically unbiased 
SMG samples.

Studying the X-ray properties of SMGs is difficult for
two main reasons.
First, the X-ray counterparts to SMGs are extremely faint.  
The count rate is so low that 
even the deepest $Chandra$ and $XMM-Newton$ spectra
of SMGs cannot resolve features that serve as sensitive 
diagnostics of the physical conditions inside galaxies, like
the Fe\,K$\alpha$ emission line.  
Fe\,K$\alpha$ emission is a ubiquitous feature in spectra of optically-selected 
AGN up to $z \simeq 3$ \citep[e.g., ][]{brus05, chau10, iwas11}, but 
the Fe\,K$\alpha$ emission properties of SMGs remain considerably 
more uncertain \citep{alex05b}.  
Second, the requirement that SMGs need radio, (sub)millimeter, 
or mid-IR counterparts capable of nailing down their positions 
in high-resolution X-ray maps can lead to concessions of 
inhomogeneously-selected samples \citep[e.g., including radio-selected galaxies; ][]{alex05b}, yielding
results that conflict with X-ray studies of purely 
submillimeter-selected SMG samples \citep{lair10, geor11}. 
To disentangle the relationship between SMGs 
and X-ray selected AGN, we need to overcome the uncertainty introduced by 
inhomogeneously selected samples,  requiring X-ray spectral 
analyses of large, flux-limited samples of 
(sub)millimeter-selected SMGs with robust counterparts.

In this work, we report on an X-ray stacking analysis of a 
sample of 38 SMGs detected in a $1.2\,\rm mm$ map of the 
Lockman Hole North (LHN), one of the fields in the
$Spitzer$ Wide-Area Infrared Extragalactic (SWIRE) Survey
 \citep{lons03}, using data from the 
 $Chandra$-SWIRE survey \citep{poll06,wilk09}.  The  
high radio counterpart identification rate of the LHN SMG sample 
\citep[93\%; ][]{lind11} is afforded by the extremely deep $20\,\rm cm$ map of 
the same field \citep{owen09}, and allows for reliable X-ray photometry.  
The sample benefits from spectroscopic 
\citep{poll06, owen09, fiol10} and optically-derived 
photometric \citep{stra10} redshifts.
Additionally, analyses of $Herschel$  
observations of the LHN \citep{magd10, rose12} have delivered reliable 
photometric redshifts and infrared luminosities for a large fraction 
of the sample by fitting far-IR photometry with 
thermal-dust spectral energy distribution (SED) models.

In \S 2, we describe the observations used 
in our analysis.  \S 3 outlines our X-ray stacking 
technique, and our method for deriving rest-frame luminosities. In \S 4, 
we compare our results to previous X-ray studies of SMGs, and discuss 
the possible origins of the Fe\,K$\alpha$ emission seen in our stacked
spectrum.  In \S 5, we present our conclusions.  In our calculations, 
we assume a $WMAP$ cosmology with 
$H_0=70\,\rm km\,s^{-1}\,Mpc^{-1}$, $\Omega_M=0.27$, and 
$\Omega_\lambda=0.73$ \citep{koma11}.  

\section{Data and Sample Selection}
\subsection{Millimeter observations \\ and stacking sample}

Our SMG sample consists of 38 of the
41 significant detections in the $1.2\,\rm mm$ map \citep{lind11}
of the LHN made using the Max Planck Millimeter Bolometer
\citep[MAMBO; ][]{krey98} array on the Institut de Radioastronomie 
Millim\'etrique $30\,\rm m$ telescope.  We exclude one source
that lacks a plausible $20\,\rm cm$ radio counterpart (L20),  
 one that has a likely X-ray counterpart (L26), and
one nearby galaxy at $z=0.044$ (L29) from the stacking
sample.  Of our final sample
of 38 galaxies, 37 (97\%) have robust $20\,\rm cm$ radio 
counterparts with a chance of spurious association 
\citep[$P$; ][]{down86} of $P<0.05$; the remaining galaxy, 
L32, has $P=0.056$.
Stacking is performed with the coordinates of the SMGs'
radio counterparts, which have a mean offset of $2.4^{\prime\prime}$
with respect to the SMG centroids.
Five of our stacking targets have positions that are not listed in
the $20\,\rm cm$ catalog of \citet{owen08} because they 
had S/N$<5.0$ (L9, L28, and L36), or they were blended 
together with nearby radio sources (L17 and L39) 
during extraction \citep{owen08,lind11}.  The sample
has a mean redshift of $\left< z \right>=2.6$ 
(see Table \ref{t-targets}).

\subsection{$Chandra$ ACIS-I Observations}

Our X-ray data are from the $3\times3$-pointing raster mosaic
of the LHN obtained with the Advanced CCD Imaging 
Spectrometer \citep[ACIS-I; ][]{weis96} on the 
$Chandra$ X-ray telescope by \citet{poll06}.  The final mosaic 
comprises nine $70\,\rm ks$ pointings arranged with 
$\sim 2^{\prime}$ overlap (see Figure \ref{f-map}).  It covers
a total area of $\simeq 0.7\,\rm deg^{2}$ and has a limiting
conventional broad band ($B_C$; 0.5-8.0$\,\rm keV$) sensitivity of 
$\sim 4\times 10^{-16}\,\rm erg\,s^{-1}\,cm^{-2}$ \citep{poll06}.
\citet{fiol09} used these same data to search for a stacked
X-ray signal among 33 $Spitzer$ $24\,\mu\rm m$-selected 
starburst galaxies, and found no significant
$0.3$--$8\,\rm keV$-band emission.

Within the sample of 41 MAMBO detections in the LHN, only L26 has a likely
X-ray counterpart (CXOSWJ104523.6+585601) in the catalog 
of \cite{wilk09}. This X-ray source has conventional broad 
band ($B_C$; 0.5-8.0$\,\rm keV$), soft 
band ($S_C$; 0.5--2.0$\,\rm keV$), and hard band 
($H_C$; 2.0--8.0$\,\rm keV$)  X-ray fluxes of 
$f_{B_C}=2.53\times 10^{-15}\,\rm erg\,s^{-1}\,cm^{-2}$, 
$f_{S_C}=1.21\times 10^{-15}\,\rm erg\,s^{-1}\,cm^{-2}$, and
$f_{H_C}=1.57\times 10^{-15}\rm\, erg\,s^{-1}\,cm^{-2}$, respectively, and
a hardness ratio of ${\rm HR} =-0.32^{+0.31}_{-0.34}$.  The hardness ratio is 
defined by ${\rm HR} = (H_C - S_C) / (H_C + S_C)$, where $H_C$ and 
$S_c$ are the counts in the $Chandra$ conventional hard and soft 
bands, respectively.

\section{Stacking Analysis}
In this section we describe our reduction of the $Chandra$ X-ray data
products and the methods used in our stacking analysis.  We use
two techniques: (1) image-based stacking in binned X-ray maps 
(\S \ref{ss-1stack}), and
(2) a photon-based spectral stacking procedure using optimized 
apertures (\S \ref{ss-2stack}).
The following subsections describe our implementation of these
two methods.

\subsection{Image-based stacking}
\label{ss-1stack}
We generate $1^{\prime\prime} \times 1^{\prime\prime}$-pixel gridded 
maps of the total counts and effective exposure time in the 
$B_C$, $S_C$, and $H_C$ energy bands using the $Chandra$ 
Interactive Analysis of Observations \citep[CIAO; ][]{frus06} script 
\texttt{fluximage}.  The characteristic energies input to \texttt{fluximage} 
to compute effective areas were $4.00\,\rm keV$,  
$1.25\,\rm keV$, and $5.00\,\rm keV$ for the $B_C$, $S_C$, and $H_C$ 
bands, respectively.  We then used the CIAO scripts \texttt{reproject\_image}
to merge the maps of each observation into one mosaic, 
and \texttt{dmimgcalc} to produce an exposure-corrected flux 
image in units of [photons $\rm \,cm^{-2}\,s^{-1}$]. 

\begin{figure}
\epsscale{1}
\plotone{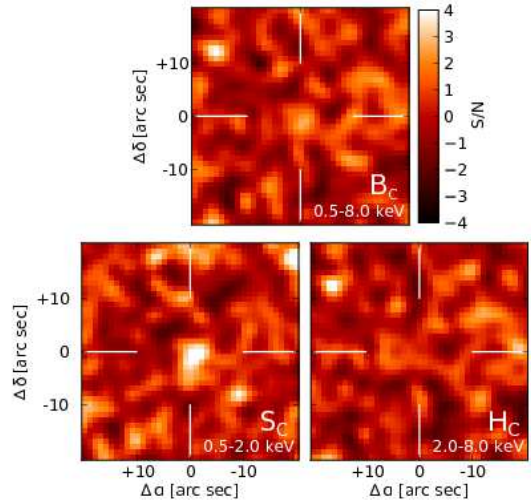}
\caption{Stacked X-ray images showing the S/N in the
$B_C$, $H_C$, and $S_C$ $Chandra$ energy bands.
The white cross hairs mark the stacking center.  The color 
stretch is S/N=[-4,+4].}
\label{f-stamp}
\end{figure}

Figure \ref{f-stamp} shows the resulting stacked image in 
each energy band.  The $40^{\prime\prime} \times 40^{\prime\prime}$ 
S/N postage stamp images are shown with a color stretch from 
S/N=$-4$ to $+4$. The peak S/N is 3.2, 4.8, and  2.0 in the 
$B_C$, $S_C$, and $H_C$ bands, respectively.  The peak of the 
strong stacked detection in the $S_C$ band has an offset from the mean
radio counterpart centroid position of $\lesssim 1^{\prime\prime}$.

\subsection{Optimized broad-band stacking}
\label{ss-2stack}

Our second stacking technique does not use a binned X-ray map.  Instead, we
compute the stacked count rate and flux by directly counting photons
at the stacking positions.  The photometric aperture 
at each stacking location is derived using a technique similar to the 
optimized stacking algorithm presented in \citet[][ supplementary 
information]{trei11}.

\begin{figure}
\epsscale{1.0}
\plotone{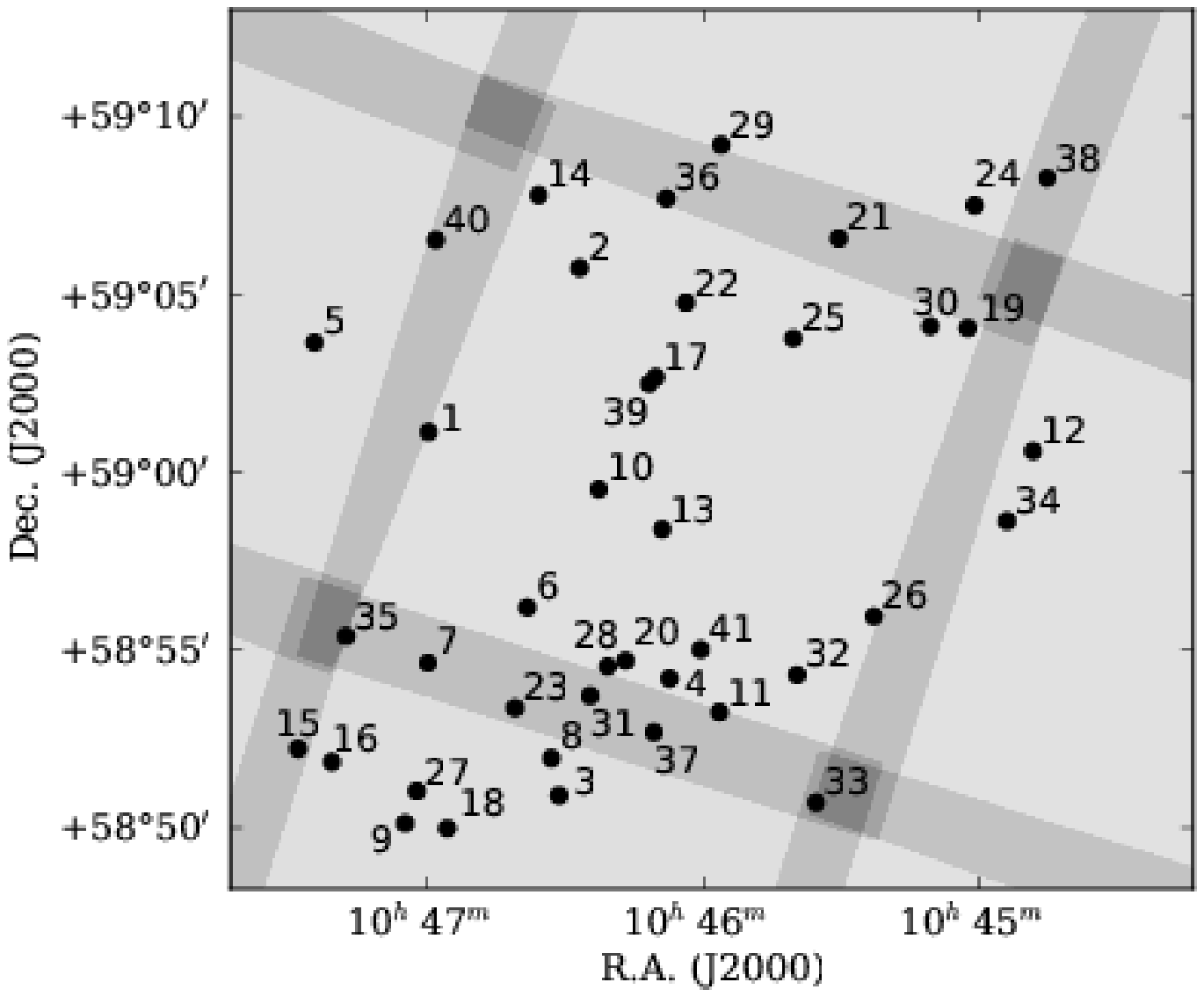}
\caption{SMG positions inside the LHN.  The filled 
circles mark the locations of the stacking targets used in
this work and are labeled according to the SMGs' 
ID numbers from Table \ref{t-targets}.  The locations of the 
source without a reliable radio counterpart (L20), the nearby 
galaxy at $z=0.044$ (L29), and the source with a likely X-ray 
counterpart (L26) are also shown even though they are not
used in our stacking analysis.  The greyscale image shows
the relative $Chandra$ effective exposure time across the field.}
\label{f-map}
\end{figure}

The size and shape of the aperture at each stacking position is chosen 
to maximize the point-source S/N at that position on the ACIS-I chips.  The 
apertures are constructed as follows.  For each stacking position 
(shown in Figure \ref{f-map}), we 
(1) use the CIAO script \texttt{mkpsf} to generate a 2D image of the 
local $Chandra$ PSF, (2) convolve this PSF with a Gaussian smoothing
kernel (see below), and (3) find the enclosed-energy fraction (EEF) 
contour $C_{\rm EEF}$ that maximizes the S/N of the flux within the 
aperture.  The area enclosed by this contour defines the aperture.  Because 
Poisson noise from the X-ray background is stronger than the flux at each 
position and the average exposure time does not change rapidly with 
increasing aperture size,  the S/N within the aperture can be parametrized 
by $C_{\rm EEF}$ as 

\begin{equation}
{\rm S/N} \propto \frac{\rm EEF}{\sqrt{A(C_{\rm EEF})}},
\label{e-aper}
\end{equation}
where  $A(C_{\rm EEF})$ is the total area enclosed by the contour $C_{\rm EEF}$.  
This expression is the same as that derived by \cite{trei11}, except that 
instead of using only circular apertures, we allow for non-axisymmetric apertures that
follow the local shape of the $Chandra$ PSF.

We compensate for the change in shape of the $Chandra$ PSF
with photon energy by generating two optimal apertures at each stacking
position, one for each energy band (i.e., for characteristic energies of
$1.25\,\rm keV$ and $5\,\rm keV$).
Parts of the LHN $Chandra$ mosaic were imaged
multiple times due to the overlapping edges of the individual exposures 
(see Figure \ref{f-map}).  For these positions we find 
the total effective aperture by maximizing Equation \ref{e-aper} 
using a linear combination of PSFs, one for each overlapping 
observation.  

We broaden the local $Chandra$ PSFs to accommodate photons
that do not lie at the stacking centers 
due to intrinsic wavelength offsets in the galaxies
and astrometric errors.  Previous X-ray stacking 
analyses find the optimal aperture radius to be 
$\sigma=1.25$--$3.0^{\prime\prime}$  \citep{lehm05,geor11}.
We find similarly that our broad band stacked
S/N is maximized with a smoothing kernel radius 
$\sigma_{\rm kernel}=1.3^{\prime\prime}$ (see Figure \ref{f-kernel}), so
we adopt this value for our subsequent broadband photometry.
The smallest angular separation of any pair of stacking targets 
($15^{\prime\prime}$ for L17 and L39) 
is larger than the  maximum radial extent of the largest X-ray stacking 
aperture, so we can ignore the effects of X-ray blending within our sample.

The photons used for background subtraction are collected from 
arrays of large circular apertures positioned next to each stacking 
target position.  The apertures are manually positioned to exclude 
any bright nearby X-ray sources that could contaminate the background
estimate.  To avoid possible systematic uncertainties associated with 
the background subtraction \citep[see, e.g., ][]{trei11, will11}, we do 
not impose any S/N-based clipping or additional filtering in
the background regions.  

Table \ref{t-data} shows the average stacked count rate and energy 
flux per galaxy in the three broad energy bands.   We find a significant
stacked detection in the soft band, and no significant detection in the
hard band.  To convert the stacked count rate into energy flux, we used the 
web-based CIAO Portable Interactive Multi-Mission 
Simulator\footnote{http://cxc.harvard.edu/toolkit/pimms.jsp} 
(PIMMS- version 4.4; Cycle 5).  The fluxes are corrected for
Galactic absorption using the column density in the direction of 
the LHN center, $N_{\rm H}=6.6\times 10^{19}\,\rm cm^{-2}$ \citep{star92}. 
Our non-detection in the hard band leaves our calculation of the
hardness ratio relatively unconstrained, $ \rm HR = -0.68^{+0.51}_{-0.32}$ 
(setting a limit on the photon index $\Gamma > 1.2$), although 
it is clear that our sample has a 
steeply declining photon spectrum
characteristic of star-forming galaxies \citep[e.g., ][]{rana12}. 
This estimate of HR was made after subtracting out the count rate in the 
soft band that is due to the strong Fe\,K$\alpha$ line 
(see \S \ref{s-spectrum}), which is a  $\sim 20\%$ contribution for 
$1.3^{\prime\prime}$-broadened photometric apertures.

A high photon index of $\Gamma=1.6$ ($\rm HR\simeq -0.37$) is 
found by \citet{lair10}, who stack on SCUBA-detected  SMGs in the 
$Chandra$ Deep Field North (CDF-N).  An even steeper photon index
of $\Gamma=1.9$ ($\rm HR\simeq -0.49$) is measured by \citet{geor11}, 
who stack on LABOCA-detected SMGs in the Extended Chandra 
Deep Field South (ECDF-S).   We have used values of
$\Gamma=1.6$ and $\Gamma=1.9$ to compute the stacked
flux in each energy band (e.g., see Table \ref{t-data}), although 
the difference between the two estimates is less than the Poisson 
uncertainty (see Table \ref{t-data}).  

\subsection{Optimized spectral stacking}
\label{s-spectrum}
In addition to stacked broad-band fluxes, we also calculate 
the observed-frame and rest-frame stacked count-rate spectra 
for our sample.  

\begin{figure}
\epsscale{0.8}
\plotone{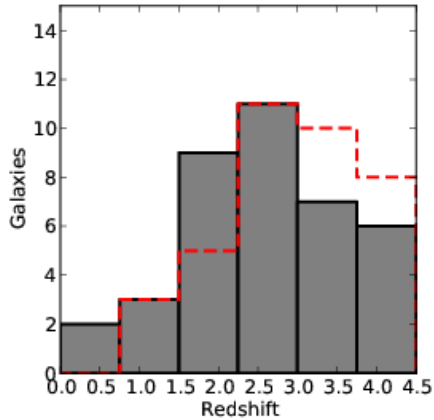}
\caption{Redshift distribution for our sample of 38 SMGs.
The filled histogram represents the redshifts used in this work, 
with $\left< z \right> = 2.6$.
The red dashed line represents the values as presented in
\citet{lind11}, which relied more heavily on spectral
index-based photometric redshifts.}
\label{f-redshifts}
\end{figure}

For each stacking target, we use redshift information in the following
order of priority, subject to availablility:
(1) spectroscopic \citep{poll06,owen09,fiol10}, 
(2) $Herschel$-based photometric 
\citep{magd10, rose12}, (3) $AA$-quality optical-based 
photometric \citep{stra10}, and (4) millimeter/radio photometric estimated using the
\citet{cari99} spectral index 
$\alpha^{20\,\rm cm}_{850\rm\,\mu m}$ technique \citep{lind11}.
The redshift distribution of our sample is shown in Figure \ref{f-redshifts}.

We use a flat sum of the observed counts at each stacking-target
position with no weighting factors.  Although this technique gives 
more weight to the brightest members of the stack, it is 
necessary given that $none$ of our stacking targets are individually 
detected and therefore S/N-based weights \citep[used in, e.g., ][]{trei11} 
cannot be reliably assigned.
For the rest-frame data, we separately coadd, blueshift, and bin 
the background photons to avoid creating artificial spectral 
features \citep[see, e.g., ][]{yaqo06}.

The uncertainty in the rest-frame energy of the photons $\Delta E_{\rm rest}$
as a function of the observed photon energies $E_{\rm obs}$ 
due to the typical redshift error $\Delta z$ is estimated by
the equation 
\begin{equation}
\Delta E_{\rm rest} = E_{\rm obs}\,\frac{\left< \Delta z\right> }{1 + \left< z \right>}.
\label{e-zerrors}
\end{equation}
This uncertainty is always larger than the energy resolution of the 
ACIS-I chips, so we set the rest-frame energy
bin widths to match $\Delta E_{\rm rest}$ (Equation \ref{e-zerrors}) 
using our sample's average redshift $\left< z \right> = 2.6$ and redshift
uncertainty $\left< \Delta z \right> = 0.4$ (see Table \ref{t-zerrors}).

\begin{figure}
\epsscale{1.0}
\plotone{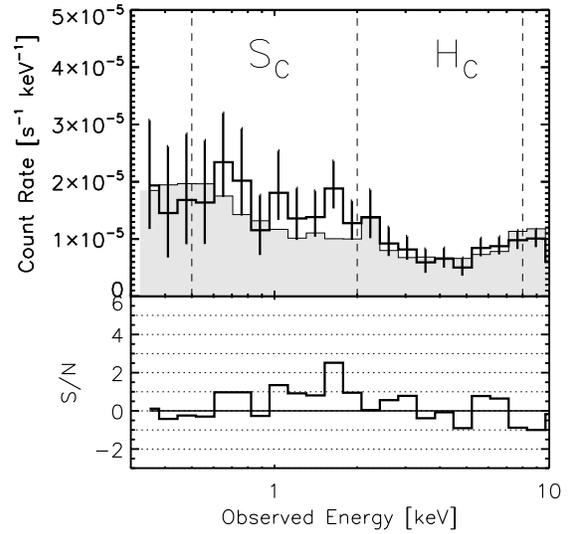}
\caption{Stacked observed-frame X-ray spectrum of our SMG
sample.  {\it Above:}  Count rate spectrum.  The solid line and 
filled region represent the stacked on-target and background 
count rates. The error bars show the Poisson 68\% double-sided 
confidence intervals due to the background.  The dashed lines
mark the extent of the $S_C$ and $H_C$ energy bands.  
{\it Below:} S/N of each bin in the above spectrum.}
\label{f-counts-obs}
\end{figure}

\begin{figure}
\epsscale{1.0}
\plotone{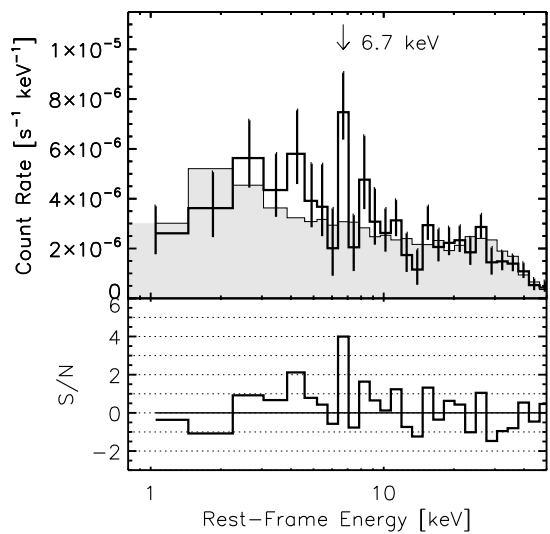}
\caption{Stacked rest-frame X-ray spectrum of our SMG
sample.  {\it Above:}  Count rate spectrum.  The solid line and 
filled region represent the stacked on-target and background 
count rates. The error bars show the Poisson 68\% double-sided 
confidence intervals due to the background only.  The arrow 
marks the location of the rest-frame bin that is centered 
at $6.7\,\rm keV$.  {\it Below:} S/N of each bin in the above
spectrum.}
\label{f-counts}
\end{figure}

Figures \ref{f-counts-obs} and \ref{f-counts} show the net 
observed and rest-frame count-rate spectra for our SMG sample, 
respectively.  
The rest-frame spectrum contains a $4\sigma$ 
emission feature with a centroid near $6.7\,\rm keV$, which we 
attribute to Fe\,K$\alpha$ line emission from a mixture of Fe 
ionization states including Fe\,XXV (see \S \ref{sss-line}).  It is apparent
from this rest-frame spectrum that a significant fraction of the 
observed soft-band flux is due to the Fe\,K$\alpha$ emission line.
If strong unresolved Fe\,K$\alpha$ emission is a 
common feature in the X-ray spectra of other SMG samples, it may 
artificially lower their measured HR values by inflating their observed
soft-band fluxes.

\begin{figure}
\epsscale{1.0}
\plotone{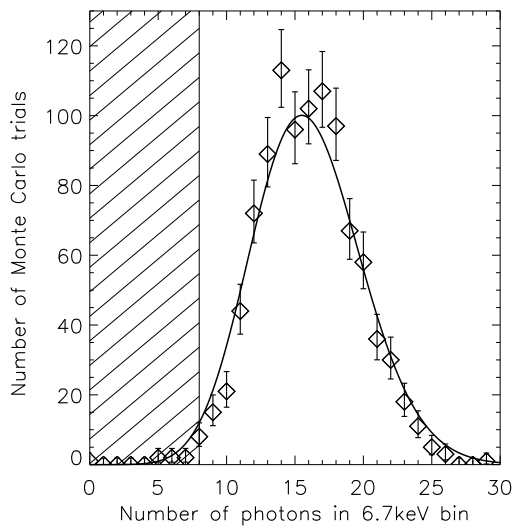}
\caption{Bootstrap-Monte Carlo analysis of the number of
Fe\,K$\alpha$ counts in the $6.7\,\rm keV$-bin of the rest-frame 
stacked spectrum (Figure \ref{f-counts}).  The histogram 
represents the total number of Monte Carlo trials returning 
the given number of total counts when randomly selecting
38 stacking positions from the original 38, with replacement.  
16 counts lie in the $6.7\,\rm keV$ bin of the real data, and the 
mean Monte Carlo result is 15.9 counts.  The solid curve shows 
an ideal Poisson distribution with mean $\mu=16$.  The shaded region 
represents the background level in this bin (8 counts).}
\label{f-boot}
\end{figure}

To ensure our stacking signal is not the result of contamination
from a few strong targets,  we performed a bootstrapping Monte Carlo 
analysis to recover the probability distribution for the 
single $6.7\,\rm keV$-energy bin ($\rm bin_{6.7}$).
The mean number of on-target counts in $\rm bin_{6.7}$ is 16, while 
the mean number of background counts in the bin is 8.  
Figure \ref{f-boot} shows that the resulting distribution closely matches 
that of an ideal Poisson distribution with a mean of 16, 
confirming that our stacking signal is characteristic of the entire sample, 
not a few outliers.

\subsection{Estimating $L_X$ and $L_{\rm Fe\,K\alpha}$}
\label{ss-feka}
The mean stacked {\it rest-frame} hard-band X-ray luminosity, 
$L_{H_C}$, of our sample is given by

\begin{equation}
\left< L_{H_C}\right> = 4\,\pi\,f_{H_C}\left< d_L^2\right>,
\end{equation}
where $f_{H_C}$ is the stacked energy flux per galaxy 
inside the observed $0.56$--$2.22\rm\,keV$ energy band 
(the $2.0\mbox{--}8\,\rm keV$ energy band, redshifted
by $\left< z \right>=2.6$), and $d_L$ is the luminosity distance
at the redshift of each stacking target.  We convert the
observed count rate to an energy flux using PIMMS.  The resulting
rest frame X-ray luminosity is 
$\left< L_{H_C}\right> = (3.0\pm1.1) \times 10^{42}\,\rm erg\,s^{-1}$.

To estimate the equivalent width and line flux of the Fe\,K$\alpha$
feature, we assume that the line emission is contained only within the 
single elevated bin at $6.7\,\rm keV$ (see Figure \ref{f-counts}) and 
estimate the local count-rate continuum around the 
$\rm Fe\,K\alpha$ feature by averaging together the 8 
bins between 3--9\,keV (excluding the bin containing the line). 
This 
results in an equivalent width of
$\rm EW=3.9\pm 2.5\,\rm keV$.  Although the EW is relatively
unconstrained, it is $>1\,\rm keV$ with 90\% confidence.
Using
the nominal equivalent width and 
Equation \ref{e-pflux}, we find a mean stacked  
$\rm Fe\,K\alpha$ line flux of
$\left< f_{\rm Fe\, K\alpha}\right> \simeq 2.1\times 10^{-17} 
\,\rm erg\,cm^{-2}\,s^{-1}$, and a mean line luminosity 
$\left< L_{\rm Fe\,K\alpha}\right> = 
(1.3 \pm 0.4) \times 10^{42}\,\rm erg\,s^{-1}$.

\begin{figure}
\epsscale{0.8}
\plotone{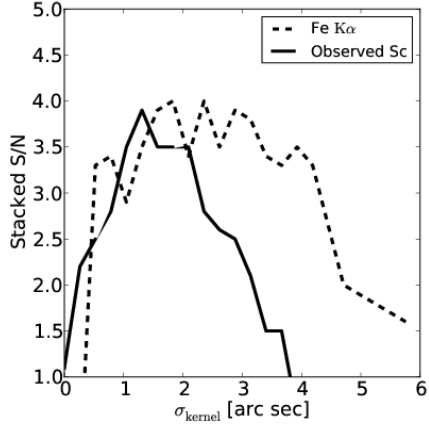}
\caption{Stacked S/N of the observed $S_C$ and rest-frame 
Fe\,K$\alpha$ line fluxes as a function of the optimal aperture
broadening kernel size $\sigma_{\rm kernel}$. }
\label{f-kernel}
\end{figure}

Figure \ref{f-kernel} also shows that that the S/N of the
Fe\,K$\alpha$ signal only drops at a larger radius 
than the broad-band signal.  If we interpret the X-ray 
continuum as originating from the galaxies' nuclear 
regions, then this relative offset between the 
Fe\,K$\alpha$ emission and the X-ray continuum 
indicates that the Fe\,K$\alpha$ photons in our sample 
are systematically offset from the galaxies' centers.  By 
measuring the distance between the peaks of the 
two curves in Figure \ref{f-kernel}, we estimate 
the radial offset to be $\sim 1^{\prime\prime}$.  For our 
computation of the Fe\,K$\alpha$ line luminosity, we adopt an 
aperture broadening kernel suited to maximize the S/N 
of the Fe\,K$\alpha$ emission line, 
$\sigma_{\rm kernel}=2.4^{\prime\prime}$.

We use a two-sample Kolmogorov-Smirnov (KS) test 
to determine the significance of the apparent 
angular extension of the Fe\,K$\alpha$ emission relative to the
continuum emission.  First, we compute the $Chandra$ PSF at the location of each 
stacking target, then sample the PSFs at the positions of their
respective collections of optimally-selected photons 
(see \S \ref{s-spectrum}).  The PSFs are peak-normalized 
and smoothed by an amount $\sigma_{\rm smooth}$ to reflect
intrinsic wavelength offsets and astrometric errors in the 
photon positions.  We then compare the cumulative distributions 
of the observed soft-band (0.5--2.0$\,\rm keV$) continuum 
photons (excluding those in the rest-frame Fe\,K$\alpha$ bin) 
and the rest-frame Fe\,K$\alpha$ photons using the 
two-sample KS test to determine with what 
confidence $1-p$ ($p$ is the KS test significance) 
we can rule out the null hypothesis that the two samples 
are drawn from a common distribution.
When using all 38 stacking positions, we find a maximum confidence of 
$1-p=0.71$ at $\sigma_{\rm smooth}=0.5^{\prime\prime}$
(63 continuum counts and 15 Fe\,K$\alpha$ counts).
When we use only the stacking positions that have $\ge1$ Fe\,K$\alpha$ 
photon, the maximum confidence occurs at the same value of 
$\sigma_{\rm smooth}$ but has a reduced $1-p=0.45$ 
(29 continuum counts and 15 Fe\,K$\alpha$ counts).
Therefore, the extension in the Fe\,K$\alpha$ emission relative to the
continuum emission indicated by Figure \ref{f-kernel} is a 
$\gtrsim 1\sigma$ (71\% confidence) effect, whose significance 
is limited primarily by the small number of Fe\,K$\alpha$ photons.

\section{Obscuration and Star Formation Rate}
\label{s-obscure}
Two galaxies in our sample have 
$L_{\rm IR}$ [8--1000$\,\rm\mu m$] estimated from \citet{magd10}, and 
12 from  \citet{rose12}. For the remaining
galaxies without SED fits, we estimate $L_{\rm IR}$
by scaling the SED from the nearby, bolometrically-star formation dominated ULIRG, $\rm Arp\,220$:

\begin{equation}
L_{\rm IR}= L_{\rm IR}^{\rm Arp220}
\left( \frac{S_{1.2\,\rm mm}}{S_{\nu_{0}}^{\rm Arp220}} \right)
\left[ \frac{d_{L}(z)}{d_{L}(z_{0})} \right]^2
\left( \frac{1+z_{0}}{1+z} \right),
\end{equation}
in terms of 
\begin{equation}
\nu_{0}= {1.2\,\rm mm} \times \left( \frac{1+z}{1+z_{0}} \right),
\end{equation}
$L_{\rm IR}^{\rm Arp\,220}=1.3\times10^{12}\,L_{\odot}$, the observed $1.2\rm mm$ 
flux density $S_{1.2\rm mm}$, luminosity distance 
$d_{L}$, target redshift $z$,
and $\rm Arp\,220$ redshift $z_{0}=0.018$.  
The $L_{\rm IR}$ values for all stacking targets are presented
in Table \ref{t-targets}; the average value 
of our whole sample is 
\begin{equation}
\left< L_{\rm IR}\right> = (2.4 \pm 0.2)\times 10^{46}\,\rm erg\, s^{-1}.
\end{equation}
The mean $20\,\rm cm$ radio luminosity density $L_{20\,\rm cm}$
is calculated using our sample's redshift distribution and
$20\,\rm cm$ flux densities \citep{owen08,lind11}:
\begin{equation}
\left< L_{\rm 20\,cm}\right> = (2.5 \pm 0.3)\times 10^{31}\,\rm erg\, s^{-1}\,Hz^{-1}.
\end{equation}

We estimate the  average star formation rate in our
sample using the scaling relations of of \citet{kenn98} in the infrared
and  \citet{bell03} at radio wavelengths, giving 
${\rm SFR}_{\rm IR}\simeq (1100\pm 100)\, M\rm_{\odot}\,yr^{-1}$ and 
${\rm SFR}_{\rm radio}\simeq (1400\pm 200)\, M\rm_{\odot}\,yr^{-1}$, respectively.  
These values are consistent with each other, but greater than the estimate
using the X-ray scaling relation from \citet{vatt11}, 
${\rm SFR}_{X}\sim (500\pm 300) \,M\rm_{\odot}\,yr^{-1}$.  
All three scaling relations assume a \citet{salp55} initial mass
function with limiting masses of 0.1 and $100\,M_{\odot}$.
The SFR estimated using the X-ray luminosity may be low due to 
intrinsic absorption.   We can derive a lower limit on
the average absorbing column in our sample by computing how much
obscuration is required to reduce the  
value of $\rm SFR_{X}$ from an intrinsic value consistent 
with $\rm SFR_{IR}$ and $\rm SFR_{radio}$.
In this case, we would require 
$N_{\rm H}\sim 2.3\times 10^{23}\,\rm cm^{-2}$ based on our observed flux
in the $0.55$--$2.77\,\rm keV$ band assuming $\Gamma=1.6$
and using $\left< z \right>=2.6$.
If we use this argument to estimate  the $unabsorbed$ X-ray luminosity, 
we find $\left< L_{H_C}\right> \simeq 9.2 \times 10^{42}\,\rm erg\,s^{-1}$.

\section{DISCUSSION}

\subsection{Comparison to previous surveys}

In this section, we compare our results to those of 
previous X-ray analyses of SMG samples from the CDFN 
\citep{alex05b,lair10} and the (E)CDF-S \citep{geor11}.

\subsubsection{Detection rate}
\label{ss-detection-rate}

With only one significant X-ray counterpart in the LHN, the 
\citet{lind11} SMG sample has an X-ray detection rate of 
$2^{+6}_{-2}\%$.  
\citet{alex05b} find 
a high X-ray detection rate of $85^{+15}_{-20}\%$ among
SMGs and submillimeter-targeted radio galaxies (which 
constitute $70\%$ of their sample) in the CDFN.  \citet{lair10} 
find a lower detection rate of $45\pm8\%$ using their
purely submillimeter-selected sample derived from the 
inhomogenously covered SCUBA supermap \citep{bory03}.
The LESS sample of \citet{geor11} is also 
purely submillimeter-selected and has an X-ray detection rate
of $11^{+4}_{-3}\%$.  However, unlike 
the SCUBA supermap, the LESS survey is produced 
with a single observing mode, and with uniform coverage.

We can place these four surveys in a common framework
if we ask what fraction of SMGs in each survey
have X-ray counterparts above the X-ray detection threshold 
in the LHN.
In this case,  we find 11 of 20 ($55^{+14}_{-13}\%$) for
\citet{alex05b},  0 of 35 ($0^{+7}_{-0}\%$) for \citet{lair10}, 
and 11 of 126 ($9^{+3}_{-3}\%$) for \citet{geor11}.  The latter
two are in agreement with our sample in the LHN.
These results indicate that a lower X-ray detection rate 
may be more characteristic of strictly submillimeter-detected SMGs
from surveys made with uniform coverage.

\subsubsection{$L_{\rm Fe\,K\alpha}$ vs. $L_{\rm IR}$ vs. $L_{20\,\rm cm}$}

\begin{figure}
\epsscale{1.0}
\plotone{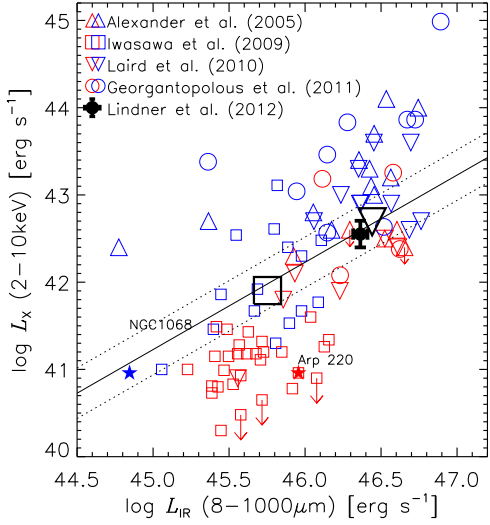}
\caption{Log X-ray luminosity vs. log infrared (8--1000$\,\rm\mu m$) 
luminosity for our SMG sample compared to other 
stacked and individually detected SMGs, LIRGs, and 
ULIRGs from the literature:  SMGs in the 
CDF-N from Alexander et al. (2005b, triangles) and Laird et al.
(2010, upside-down triangles), SMGs in the 
ECDFS from Georgantopoulos et al. (2011, circles),  
and nearby LIRGs and ULIRGs from Iwasawa et al. (2009, squares). 
Red and blue symbols represent galaxies with  
``star formation-only'' or ``AGN-required" X-ray 
spectral classifications, respectively. 
The line represents
the relation for star-forming galaxies in the local
Universe \citep{rana03}.  The black hollow symbols
represent the average luminosities of 
all LIRGs/ULIRGs (Iwasawa et al. 2009, square), and 
all SMGs (including the stacked contribution) of the 
SCUBA supermap detections in the CDFN 
(Laird et al. 2010, upside-down triangle).}
\label{f-lirlx}
\end{figure}

Figure \ref{f-lirlx} shows our sample's average X-ray (corrected only
for Galactic absorption), radio, and IR luminosities compared to those of 
other stacked SMG samples \citep{lair10,geor11}, individually 
X-ray-detected SMGs \citep{alex05b, lair10, geor11}, and nearby 
LIRGs and ULIRGs \citep[sample drawn from ][]{iwas09}.  
Where available, we use the $L_{\rm IR}$ value from Table A2 of 
\citet{pope06} for the SMGs from the CDFN.  For the twelve 
SMGs in \citet{alex05b} that are not in the catalog 
of \citet{pope06}, we scale $L_{\rm \rm FIR}\longrightarrow L_{\rm IR}$
using the average conversion factor $f$ for the eight SMGs common
between the two samples, $f=1.42$.  The $870\,\rm\mu m$-detected SMGs 
from the (E)CDF-S are plotted with $L_{\rm IR}=10$--$1000\,\rm\mu m$.
The local LIRGs and ULIRGs from \citet{iwas09} also have
their $L_{\rm FIR} (40$--$400\,\rm\mu m)$ estimates scaled by $f=1.42$.
We also show the total sample luminosity average for \citet{lair10}, 
including the contribution from their stacked SMGs that were not
individually detected in the X-ray. 
The average properties of our 
stacking sample are in agreement with the total 
luminosities of \citet{lair10}.

\begin{figure}
\epsscale{1.0}
\plotone{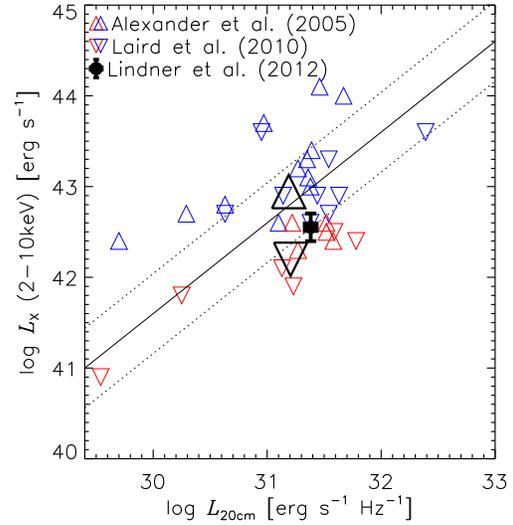}
\caption{X-ray  luminosity versus $20\,\rm cm$ spectral power.
The symbols and colors are the same as in Figure \ref{f-lirlx}.
The over-plotted solid and dashed lines show the correlation
between $L_X$ and $L_{20\,\rm cm}$ along with $1\sigma$ 
errors for star-forming galaxies \citep{vatt11}.}
\label{f-l20lx}
\end{figure}

Figures \ref{f-lirlx} and \ref{f-l20lx} also indicate the 
AGN classification of each galaxy.  Galaxies whose 
mid-IR or X-ray spectral 
properties are consistent with emission produced entirely by 
star formation are plotted in red, while those requiring the
presence of an AGN are shown in blue.
\citet{geor11} divide their sample by using a probabilistic 
approach;  those galaxies requiring the presence of a torus-dust 
component in their mid-IR SED according to an F-test are 
categorized as AGN.  \citet{lair10} and \citet{alex05b} separate out the AGN
based on the most favored model of their X-ray spectra 
according to the \cite{cash79}
statistic.  The sample of \citet{iwas09} is divided based on
hardness ratio.

The division between AGN and non-AGN systems 
can be roughly determined based on the X-ray scaling relations 
of purely star-forming galaxies in the local universe 
\citep{rana03,vatt11}, shown as solid black lines.  The 
average properties of our stacking sample lie very 
near the \citet{rana03} relation.  Considering
the substantial intrinsic scatter in the spectral
classifications of the \citet{lair10} sample, 
our stacking sample also probably contains a 
substantial fraction of both 
star-formation-only and AGN-required systems.

\subsubsection{Fe K$\alpha$ emission properties}
\label{sss-line}

The Fe\,K$\alpha$ photons in our stacking sample 
may be more spatially extended than the continuum photons 
by $\sim 1^{\prime\prime}$ (see \S \ref{ss-feka}). 
Extended and misaligned 
Fe\,K$\alpha$ emission has been observed in 
Arp\,220 \citep{iwas05} and NGC\,1068 \citep{youn01}.
We may also be blending together the emission from 
multiple components of merging systems of which
only one component has strong Fe\,K$\alpha$
emission \citep[like, e.g., Arp 299; ][]{ball04}.

\begin{figure}
\epsscale{1.0}
\plotone{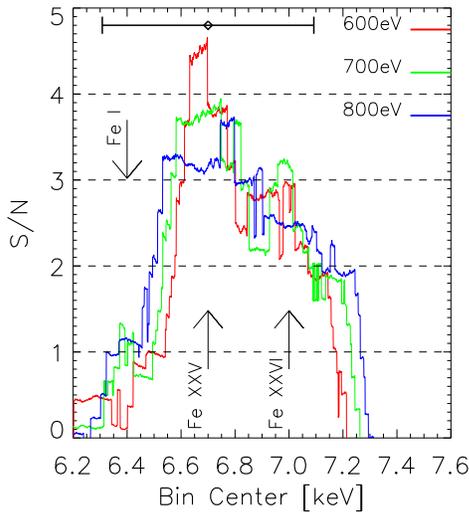}
\caption{S/N of the Fe\,K$\alpha$ emission as a function of bin position
and width.  The red, green, and blue lines use bin widths of $600\,\rm eV$, 
$700\,\rm eV$, and $800\,\rm eV$ (the bin width used in the rest frame 
spectrum in Figure \ref{f-counts} is $698\,\rm eV$ and is determined by 
the typical redshift uncertainty in our sample).  The black arrows mark  
the energies of K$\alpha$ photons from Fe\,I (6.4\,\rm keV), 
Fe\,XXV (6.7\,\rm keV) and Fe\,XXVI (7.0\,\rm keV).  The horizontal error
bar shows the formal uncertainty in the line centroid 
$\sigma_{\rm centroid}=\pm 398\,\rm eV$ centered on $6.7\,\rm keV$.}
\label{f-ion}
\end{figure}

The bin width in our 
stacked rest-frame X-ray spectrum, which is
set by the redshift uncertainties of our SMG sample,
 is larger than the rest energy separation between 
Fe\,K$\alpha$ emission from neutral and 
highly-ionized iron ($0.3\,\rm keV$); 
therefore, it is difficult to determine the average 
Fe ionization fraction in our sample.  
Close inspection of the photons near the rest-frame 
Fe\,K$\alpha$ line (see Figure \ref{f-ion}) reveals 
a range of values between 
$6.4\,\rm keV$--$7.2\,\rm keV$, with
a local maximum at $6.7\,\rm keV$. 
Given that the Fe line photons are contributed fairly evenly by the 38 
targets in our stacking sample, and have been assigned to
their bins based on a wide variety of redshift estimation techniques (spectroscopic, optical-photometric, $Herschel$-photometric, 
millimeter/radio-photometric), they are unlikely all to be 
systematically biased high or low.  Therefore, a significant fraction 
of the detected Fe\,K$\alpha$ photons likely originate from the highly-ionized 
species of Fe\,XXV or Fe\,XXVI.
However, the 
$\sim 10\%$ rest-frame uncertainty in the energy of each
photon implies an uncertainty in the centroid of the line
profile of 
$\sigma_{\rm centroid}=  \rm (FWHM)/(SNR) \simeq 398\,\rm eV$, 
insufficient to determine the relative fractions of each ionization 
state with certainty.

Strong emission ($\rm EW=1.8\pm0.9$) from highly ionized 
Fe\,K$\alpha$ has been observed in the nearby 
ULIRG Arp\,220 by \citet{iwas05} using $XMM-Newton$.  
\citet{iwas09} also find strong $6.7\,\rm keV$ emission 
($\rm EW=0.9\pm0.3\,\rm keV$) from the stacked spectrum of 
nearby ULIRGs (including Arp\,220) that have no evidence 
of AGN emission (termed X-ray-quiet ULIRGs).  \citet{alex05b} 
detected strong ($\rm EW\simeq 1\,\rm keV$) Fe\,K\,$\alpha$ emission in the 
stacked SMG spectrum of the six SMGs in their sample with 
$N_{H}>5\times 10^{23}\,\rm cm^{-2}$ , and find that the line centroid is 
between  $6.7\,\rm keV$ and $6.4\,\rm keV$, indicating a 
substantial contribution from highly ionized gas.

\begin{figure}
\epsscale{0.8}
\plotone{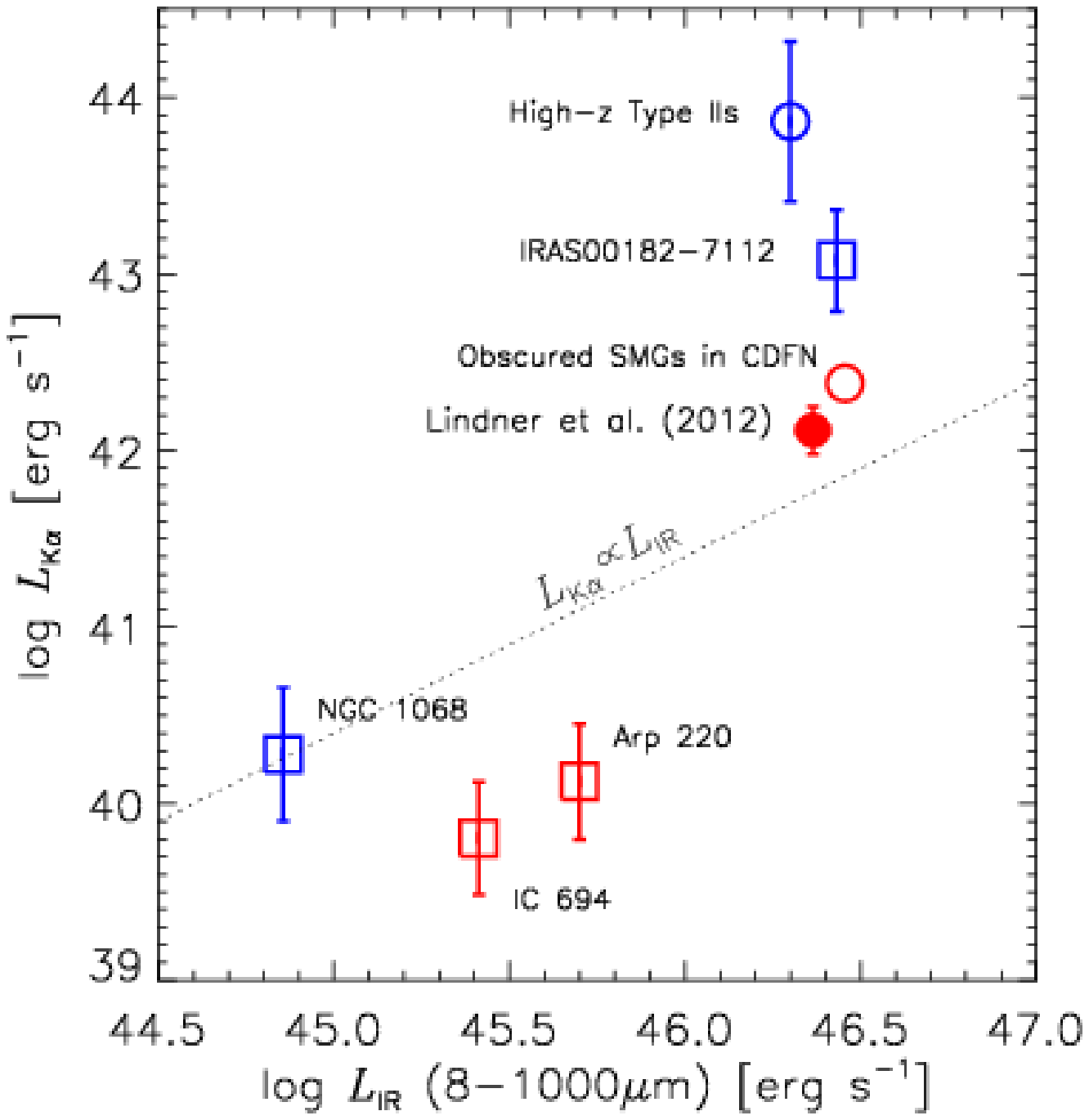}
\caption{Fe\,K$\alpha$ line luminosity $L_{\rm K\alpha}$ versus $L_{\rm IR}$ for our results
(filled red circle), stacked SMGs in the CDF-N \citep[empty red circle; ][]{alex05b}, 
and stacked high-$z$ Type II AGN \citep[empty blue circle; ][]{iwas11}.  The squares represent
the LIRG NGC\,1068 \citep{youn01}, the HyLIRG IRAS\,00182-7112 \citep{nand07}, and the
merger component IC\,694 \citep{ball04}. The dashed line represents a linear relation that is
normalized to NGC\,1068.  Blue (red) symbols denote samples/systems that do (do not) show
 strong AGN bolometric signatures. }
\label{f-lirlka}
\end{figure}

In Figure \ref{f-lirlka} we compare the relation 
between $L_{\rm K\alpha}$ and $L_{\rm IR}$ in our
sample with those for other individual systems and
stacked samples with measured
Fe\,K$\alpha$ line luminosities and 
bolometrically dominant energy sources that are
well understood.  The dashed line represents
a linear slope between $L_{\rm K\alpha}$ and $L_{\rm IR}$
and has been normalized to NGC\,1068, a
nearby prototypical Seyfert II LIRG.  Red 
symbols represent systems that
do not have significant observed AGN
bolometric contributions, like SMGs and local
X-ray quiet ULIRGs;  the blue symbols represent
systems that have significant bolometric AGN
contributions.  Figure \ref{f-lirlka} shows that 
the relative Fe\,K$\alpha$/infrared luminosity 
fraction, $L_{\rm K\alpha}/L_{\rm IR}$, increases with 
increasing $L_{\rm IR}$.  
If the Fe\,K$\alpha$ emission
is due to AGN activity, then this result may be in agreement
with the observed trend that LIRGs/ULIRGs tend
to be increasingly AGN-dominated with
increasing $L_{\rm IR}$ \citep[e.g., ][]{tran01}.

\subsection{Origin of the Fe\,K$\alpha$ emission}

This section discusses three possible
physical origins for the Fe\,K$\alpha$ emission detected
in our stacked SMG sample: supernovae, galactic-scale winds, and
AGNs.  Because a significant fraction of our sample's Fe\,K$\alpha$ 
emission likely originates from the highly-ionized species Fe\,XXV (see, e.g., 
Figure \ref{f-ion}) and because evidence for 
highly-ionized Fe\,K$\alpha$ emission from other (U)LIRGs exists at both 
high \citep{alex05b} and low \citep[e.g., ][]{iwas05} redshifts, the following
sections focus on the origin of this high-ionization component.

\subsubsection{Supernovae}

Here we consider if the observed Fe\,K$\alpha$ feature can 
be attributed to X-ray luminous supernovae.  X-ray 
observations of the supernova SN\,1986J in the nearby
spiral galaxy NGC\,831 reveal strong 
hard-band emission and a significant 
$6.7\,\rm keV$ ($\rm EW \lesssim 500\,\rm eV$) line \citep{houc98}.  
Supernova 1986J decayed in the 2--$10\,\rm keV$ band as $\sim t^{-2}$ 
from 1991 to 1996.  We will take a conservative approach and 
use only the luminosity information in this time interval 
for our calculation.  Given the X-ray luminosity and decay 
rate of SN\,1986J \citep{houc98}, and assuming the star 
formation rate of our sample of order ${\rm SFR}=10^{3}\, M_{\odot}\,\rm yr^{-1}$ 
giving a supernova rate of $10\,\rm SN\,yr^{-1}$, we would 
expect $\sim 50$ X-ray luminous supernovae to be visible at any given 
time.  The combined supernova X-ray luminosity is therefore  
$L_{H_C, {\rm SNR}}\simeq 10^{42}\,\rm erg\,s^{-1}$.  Considering the 
fact that prior to 1991 SN\,1986J was probably still dimming 
at a rate close to $\propto t^{-2}$, this calculation is an 
underestimate.  Therefore, supernovae like 1986J can satisfy 
the bolometric requirements for explaining the hard X-ray 
emission $and$ the Fe\,K$\alpha$ line that we see in our stacked SMG
sample.

However, if the supernovae associated with massive star formation are 
visible, then so must be high-mass X-ray binaries given the
short time lifetimes of massive stars.  These 
systems would dominate the hard X-ray emission from star-forming 
regions, and would severely dilute 
the Fe\,K$\alpha$ emission \citep[see, e.g., ][]{iwas09}.
We therefore rule out X-ray luminous supernovae and supernova 
remnants as the source of the Fe\,K emission in our sample.

\subsubsection{Galactic-scale winds}

As discussed in \citet{iwas05}, who consider the 
$6.7\,\rm keV$ emission line in Arp\,220, a starburst-driven 
galactic-scale superwind of hot gas is energetically plausible 
as the source of the Fe\,K$\alpha$ emission.  Large 
outflows could also explain why the the Fe\,K$\alpha$ line 
emission appears more extended than
the X-ray continuum emission in our stacking sample.  
To explore this scenario, we used the X-ray spectral-fitting 
package XSPEC \citep{arna96} to model 
an absorbed diffuse thermal X-ray (\texttt{zphabs * mekal}) spectrum 
and estimate the gas metallicity required to produce the strong high-ionization 
Fe\,K$\alpha$ emission detected in our SMG sample.  We 
computed the model EW values using the spectral window 
$6.35$--$7.05\,\rm keV$, the same energy width as the bin 
containing the Fe\,K$\alpha$ emission in our stacked rest-frame 
spectrum (Figure \ref{f-counts}), which includes all Fe\,K$\alpha$
ionization states.  We fixed the gas temperature to Arp\,220's best-fit value 
$kT=7.4\,\rm keV$ \citep{iwas05}, the gas 
density to $n=1\,\rm cm^{-3}$, the redshift to our sample's average 
$\left< z \right> =2.6$, and the obscuring hydrogen column density to 
$2.3\times10^{23}\,\rm cm^{-2}$ (\S \ref{s-obscure}).  Both the 
Fe\,K$\alpha$ line luminosity and the continuum intensity vary 
linearly with $Z$, allowing us to express the relation between 
EW and $Z$ as
\begin{equation}
{\rm EW} = \frac{6.67}{1+\frac{Z^{\prime}}{Z}}\,\rm keV,
\end{equation}
where $Z^{\prime}=5.29\,Z_{\odot}$.  EW is approximately proportional to $Z$ 
for $Z\ll Z^{\prime}$ and approaches the constant value   
$6.67\,{\rm keV}$ for $Z\gg Z^{\prime}$.
Because of this non-linear behavior, an abundance of 
$0.94\, Z_{\odot}$ can produce $\rm EW= \,\rm 1keV$ 
(90\% confidence lower-limit) while a significantly greater abundance 
$Z\simeq 7.5\,Z_{\odot}$ is needed to explain our nominal
value $\rm EW\simeq 3.9\,\rm keV$.  If a significant amount of 
our rest-frame 2--10\,keV luminosity is from X-ray binaries, incapable of 
generating the observed line emission, then the required metallicity 
would be even higher.  
While the lower limit on our measured EW can be
explained by thermal emission from a diffuse ionized plasma, especially 
considering the extreme enrichment taking place in 
systems like SMGs, generating an EW with a value close to 
our nominal measurement would require an unrealistic 
degree of high-$z$ enrichment.

\subsubsection{AGN activity}

AGNs hidden behind large hydrogen column densities
may be responsible for the observed Fe\,K$\alpha$ emission
in our sample.  The Fe\,K$\alpha$ emission line is 
the signature spectral feature of the reprocessed 
(reflected) spectrum of an AGN \citep{matt00}.  
As the ionizing luminosity increases, so does the 
ionization fraction of the gas, shifting the dominant
emission feature from $6.4\,\rm keV$ (neutral and 
intermediate ionization states) to $6.7\,\rm keV$ 
(helium-like Fe XXV) and $6.9\,\rm keV$ (hydrogen-like
Fe XXVI).

Some insight into the properties of SMGs can be gained 
from reviewing the well-studied Fe\,K$\alpha$ emission 
properties of AGN and nearby ULIRGs.  
Strong ($\rm EW\simeq 1\,\rm keV$) $6.7\,\rm keV$ emission 
has been observed in systems that
are bolometrically AGN-dominated, like 
IRAS\,00182-7112 \citep{nand07} and NGC\,1068 
\citep{youn01}, as well as systems that appear to
be energetically AGN-free, like Arp\,220 \citep{iwas05}
and IC\,694 \citep{ball04}.
However, direct evidence of a black hole accretion disk has been
observed in Arp 220 by \citet{down07} with the detection of
a compact ($0.19^{\prime\prime}\times 0.13^{\prime\prime}$) 
$1.3\,\rm mm$ continuum source in the center of the west nucleus
torus.  This source has a surface luminosity 
of $\sim 5\times 10^{14}\,L_{\odot}\,\rm kpc^{-2}$, which is 
energetically incompatible with being powered by even the 
most extreme compact starbursts known.  Only an 
accretion disk can be responsible for heating the dust.
Highly ionized Fe\,K$\alpha$ emission has
also been observed in the AGN systems 
Mrk 273 \citep{bale05}, NGC\,4945 \citep{done03}, and
NGC\,6240  \citep{boll03}, along with a neutral 
Fe\,K$\alpha$ component.

The narrow $6.4\,\rm keV$ ``cold'' 
Fe\,K$\alpha$ emission line is a ubiquitous feature in 
the spectra of optically-selected active galaxies out to high redshift 
\citep[e.g., ][]{corr08, iwas11, falo11}.  However, \citet{iwas11} 
also find evidence for highly ionized Fe\,K$\alpha$ 
emission in two subsets of their X-ray selected AGN
sample:  Type I AGN 
with the highest Eddington ratios, and 
Type II AGN with the highest redshifts.  The 
subsamples with highly ionized K$\alpha$ 
emission show no evidence of a broad line 
Fe\,K$\alpha$ feature; therefore, the highly-ionized 
Fe\,K$\alpha$ photons probably do not originate 
from the accretion disk, but from more 
distant and tenuous outflowing gas.
This scenario may also explain why the Fe K$\alpha$ photons
in our sample appear spatially extended with respect
to the X-ray continuum photons.

A significant caveat is that it remains difficult to reconcile the power source
required to produce offsets as large as $1^{\prime\prime}$ 
in the photon distribution of our sample's  
stacked Fe\,K$\alpha$ emission (relative to the nuclear continuum; see \S \ref{ss-feka}) given the sample's low average 
X-ray luminosity.  For example, 
we can calculate the maximum radial distance out to which
low-density gas can remain highly photoionized by a single ionizing source 
by assuming that our sample's stacked infrared luminosity is 
produced by deeply-buried AGNs, i.e., 
$L_{\rm ion}\simeq L_{\rm IR}=2.4\times 10^{46}\,\rm erg\,s^{-1}$. 
Using the ionization parameter $\xi \equiv L_{\rm ion}/nR^2$ 
\citep[$\log\xi\ge2.8$ is required for a significant Fe\,XXV 
ionization fraction: ][]{kall04} with $n=1\,\rm cm^{-3}$, we find 
$R_{\rm max}=\sqrt{L_{\rm IR}/n\xi}\simeq 2.0\,\rm kpc$.
At our sample's average redshift of $\left< z \right>=2.6$, 
this corresponds to a typical angular offset of 
$0.24^{\prime\prime}$.  Angular offsets 
larger than $0.24^{\prime\prime}$, like those tentatively indicated by our
sample (see Figure \ref{f-kernel}), can be explained by SMGs that host 
multiple distributed ionizing sources.  
In particular, the radio continuum emission (defining our stacking positions)
might be more closely associated with the X-ray continuum than with Fe\,K$\alpha$
line emission in a complex, multi-component system.
These results 
highlight the importance of resolving the sizes and morphologies of 
SMGs with high-resolution (sub)millimeter imaging \citep[e.g., ][]{tacc06}.

If the high-ionization Fe\,K$\alpha$ emission is ultimately 
due to star-formation processes (shocked gas from SNe), 
and the SFR is traced by the $L_{\rm IR}$, then
we should expect a linear relation between $L_{\rm K\alpha}$ and 
$L_{\rm IR}$.  If the systems with the highest 
$L_{\rm IR}$ have an infrared contribution from obscured
AGN that are not also emitting Fe\,K$\alpha$ photons, 
then we would expect a slope that is even less than
unity.  However, Figure \ref{f-lirlka} shows
that $L_{\rm K\alpha}$ is relatively much more
dominant in SMGs and high-$z$ AGN than in their 
lower-luminosity, lower-redshift analogues.  This distinction
indicates that highly ionized Fe\,K$\alpha$ emission 
cannot be explained solely by star-formation processes
and is more likely to be the result of AGN activity.

\section{Conclusions}

We analyze the X-ray properties of a complete
sample of SMGs with radio counterparts from
the LHN.  This sample's X-ray detection rate of 
$2^{+6}_{-2}\%$ is consistent with those for other 
uniformly-mapped, submillimeter-detected samples, considering 
the depth of our X-ray data.
The X-ray undetected SMGs show a strong stacked detection
in the $S_C$ band, and no significant detection 
in the $H_C$ band, similar to results from SMG 
stacking in the CDF-N \citep{lair10} and 
CDF-S \citep{geor11}.

We also use the available redshift information of our 
SMGs to compute the rest-frame, stacked count-rate 
spectrum of our sample.  The rest-frame spectrum 
shows strong ($ \rm EW> 1\,\rm keV$) emission 
from Fe\,K$\alpha$, possibly with contributions 
from Fe\,XXV and Fe\,XXVI.
A comparison with other high-ionization Fe\,K$\alpha$-emitting 
systems from the literature indicates that accretion
onto obscured AGNs is the likely explanation for 
the strong Fe\,K$\alpha$ emission line.
In our sample, the Fe\,K$\alpha$ emission
is responsible for $\sim 20\%$ of the observed
soft-band X-ray flux.  Therefore, if strong Fe line emission
is a common feature in other SMG samples, it 
would significantly decrease the measured 
values of HR and lead to overestimates 
of the continuum spectral index $\Gamma$.

We find a tentative indication (71\% confidence) that 
our sample's stacked distribution of Fe\,K$\alpha$ 
photons is more spatially extended than that of the 
X-ray continuum.  If confirmed by future studies, this 
result can help determine the physical origin of the 
prominent Fe\,K$\alpha$ emission in SMGs.

\acknowledgments
We thank Jack Hughes, Dieter Lutz, and Alain Omont for useful discussions.  
We also thank the referee for useful comments that have improved
the quality of this manuscript.
This work has been supported by NSF grant AST-0708653, and has made use 
of the NASA/IPAC Extragalactic Database (NED) which is operated by the Jet Propulsion 
Laboratory, California Institute of Technology, under contract with the National
Aeronautics and Space Administration.

{\it Facilities:} \facility{IRAM:30m}, \facility{CXO}.

\appendix
\section{Detailed descriptions of calculations}
\subsection{Optimized spectral stacking method}
\label{a-rate}
We begin by labeling all the photons within the optimized apertures 
(see Section \ref{ss-1stack}) of all of the $N=38$ stacking targets 
with the index $i$, and those within the background regions for the 
$N$ targets $i'$.  $E_i$ is the energy of the $i^{th}$ photon, and
$T_i$ is the total effective exposure time in the mosaic at the position
of $i$.  The notation $i\in j$ refers to all the photons that have  
energies located the $j^{th}$ energy bin.  The stacked mean count 
rate in the $j^{th}$ energy bin, $R_{j}$, is then

\begin{equation}
  R_j = \frac{1}{N} \sum_{i \in j} \frac{1}{T_i\,\kappa_i}
\end{equation}
where $\kappa_i$ is the aperture correction for the optimal
aperture of the energy band of the $i^{th}$ photon.  The background 
mean count rate in the $j^{th}$ bin is
\begin{equation}
  R'_j = \frac{1}{N}  
  \sum_{i' \in j} 
  \frac{1}{T_{i'}\,\kappa_{i'}\,c_{i'}},
\end{equation}
where $c_{i'}$ is the ratio of the areas of the background region 
of the stacking position of the $i'^{th}$ photon, and of the optimal 
aperture of that stacking position.  It follows that the 
expected number of background counts in the $j^{th}$ bin, 
$\tilde{N}_j$, is

\begin{equation}
\eta_j=\left< T_i\right>_{i \in j} \times R'_j.
\end{equation}

We use the double-sided 68\%-confidence upper and lower 
limits, $\eta_{j}^{\rm high}$, and  $\eta_{j}^{\rm low}$ \citep{gehr86}, 
to compute the $1\sigma$ count rate deviations in the $j^{th}$ bin 
due to the background, $\sigma^{\rm hi/low}$:

\begin{equation}
\sigma{j}^{\rm hi/low} =   
\frac{|\eta_j^{\rm hi/low} - \eta_j|}
{\left< T_{i\in j}\right>_i }
\end{equation}

Therefore, the net count rate density per galaxy in the $j^{th}$ bin, 
{$\mathbf{R_{j}}$}, is

\begin{equation}
\mathbf{R_j} = \frac{(R_j - \tilde{R_j})}{\Delta E_j}
^{+\sigma (R)_{j}^{\rm hi}/\Delta E_j}_
{-\sigma (R)_{j}^{\rm low}/\Delta E_j} \,\,\, [\,\rm s^{-1}\,keV^{-1}\,],
\end{equation}
where $\Delta E_{j}$ is the width of the $j^{th}$ energy bin.  We 
calculate the corresponding rest-frame spectrum $\mathbf{R^{\rm rest}_j}$ 
by binning the photons according to their rest-frame energies, 
$E^{\rm rest}_i=E_i\,(1+z_i)$, where $z_i$ is the redshift of the 
stacking target associated with the photon.

\subsection{Fe K$\alpha$ energy flux}

We use the CIAO script \texttt{eff2evt} to 
tabulate the local effective area $A_i$ ($A_{i'}$),
 and quantum efficiency $Q_i$ ($Q_{i'}$),  for each photon 
in the on-target (background) apertures.

The mean stacked on-target and background photon fluxes in 
the $j^{th}$ bin, $F_j$, and $F'_j$, are then
\begin{equation}
F_j = \frac{1}{N} \sum_{i \in j} 
\frac{1}{T_i\,A_i\,Q_i\,\kappa_i},
\end{equation}
and
\begin{equation}
F'_j = \frac{1}{N} \sum_{i' \in j} 
\frac{1}{T_{i'}\,A_{i'}\,Q_{i'}\,\kappa_{i'}\,c_{i'}},
\label{e-pflux}
\end{equation}

respectively.


\clearpage

\begin{deluxetable}{ccccccccl}
\tabletypesize{\scriptsize}
\tablecaption{LHN MAMBO SMGs}
\tablehead{ID & Name & RA\tablenotemark{a} & Dec\tablenotemark{a} & $z$ & $z$ type\tablenotemark{b} & $z$ reference\tablenotemark{c} &
${\rm log}\left( L_{\rm IR}/L_{\odot} \right)$\tablenotemark{d} & $L_{\rm IR}$ reference\tablenotemark{c} \\ 
& & J2000 &J2000 & &  & & &}
\startdata
L1   & $\rm MM\,J104700.1+590109$ & 161.75083 & 59.018778 & 2.562   & S       & P06 & 13.3 & R12   \\
L2   & $\rm MM\,J104627.1+590546$ & 161.61192 & 59.095778 & 4.09     & P-IR  & R12 & 12.9 & L12   \\
L3   & $\rm MM\,J104631.4+585056$ & 161.63112 & 58.848889 & 1.8       & CY     & L11 &13.0   & L12  \\
L4   & $\rm MM\,J104607.4+585413$ & 161.53050 & 58.903889 & 4.4       & CY     & L11 & 12.7  & L12  \\
L5   & $\rm MM\,J104725.2+590339$ & 161.85592 & 59.060139 & 3.00     & P-O   & S10 & 13.1 & R12    \\
L6   & $\rm MM\,J104638.4+585613$ & 161.66112 & 58.936806 & 2.03     & S       & F10 &  12.9 & R12     \\
L7   & $\rm MM\,J104700.1+585439$ & 161.75054 & 58.911500 & 2.74     & P-IR  & R12 & 12.7  & L12    \\
L8   & $\rm MM\,J104633.1+585159$ & 161.63779 & 58.866417 & 2.07     & P-IR  & R12 & 12.8    & L12 \\
L9   & $\rm MM\,J104704.9+585008$ & 161.77071 & 58.835750 & 3.9       & CY     & L11 & 12.9     & L12 \\
L10 & $\rm MM\,J104622.9+585933$ & 161.59604 & 58.993000 & 3.01     & P-IR  & R12 & 12.7   & L12 \\
L11 & $\rm MM\,J104556.5+585317$ & 161.48721 & 58.888556 & 1.95     & S       & F10 & 12.9 & R12      \\
L12 & $\rm MM\,J104448.0+590036$ & 161.19829 & 59.009972 & 2.16     & P-O   & S10 & 12.7 & R12     \\
L13 & $\rm MM\,J104609.0+585826$ & 161.53646 & 58.974583 & 0.32     & P-IR  & R12 & 12.3   & L12  \\
L14 & $\rm MM\,J104636.1+590749$ & 161.64937 & 59.130139 & 2.26     & P-IR  & R12 & 12.9 & R12    \\
L15 & $\rm MM\,J104728.3+585213$ & 161.86654 & 58.870583 & 2.78     & P-IR  & R12 & 13.3 & R12    \\
L16 & $\rm MM\,J104720.9+585151$ & 161.83633 & 58.864722 & 3.24     & P-IR  & R12 & 12.8   & L12  \\
L17 & $\rm MM\,J104610.4+590242$ & 161.54383 & 59.045000 & 2.65     & P-IR  & R12 & 12.7   & L12  \\
L18 & $\rm MM\,J104655.7+585000$ & 161.73021 & 58.834444 & 2.15     & P-IR  & R12 & 12.9   & L12  \\
L19 & $\rm MM\,J104502.1+590404$ & 161.25833 & 59.067583 & 4.1        & CY   & L11 & 12.6    &L12  \\
L20 & $\rm MM\,J104617.0+585444$ & 161.57058 &  58.913694 & $>4.6$ & CY   & L11&     --      & --                  \\
L21 & $\rm MM\,J104530.3+590636$ & 161.37575 & 59.110083 & 2.52       & P-IR & R12 & 12.6   & L12   \\
L22 & $\rm MM\,J104603.8+590448$ & 161.51763 & 59.080389 & 1.44       & P-IR & R12 & 12.3 & R12     \\
L23 & $\rm MM\,J104641.0+585324$ & 161.67129 & 58.890500 & 3.6         & CY    & L11 & 12.6  & L12     \\
L24 & $\rm MM\,J104500.5+590731$ & 161.25279 & 59.126000 & 3.24       & P-O   & S10 & 12.7 & R12      \\
L25 & $\rm MM\,J104540.3+590347$ & 161.41679 & 59.063333 & 3.5         & CY    & L11 & 12.6   & L12    \\
L26 & $\rm MM\,J104522.8+585558$ & 161.34833 & 58.933611  & $ >5.0$ &    CY & L11 &     --     &  --                 \\
L27 & $\rm MM\,J104702.4+585102$ & 161.76000 & 58.850861 & 1.62       & P-IR  & R12 & 12.7   & L12   \\
L28 & $\rm MM\,J104620.9+585434$ & 161.58804 & 58.909722 & 3.8         & CY    & L11 & 12.5    & L12   \\
L29 & $\rm MM\,J104556.1+590914$ & 161.48129  & 59.154528 &  0.044    &   S    & O09&  8.0& R12           \\
L30 & $\rm MM\,J104510.3+590408$ & 161.29329 & 59.068639 & 0.71        & P-IR  &  R12 &12.5   & L12   \\
L31 & $\rm MM\,J104624.7+585344$ & 161.60379 & 58.896444 & 2.90        & P-IR  & R12 & 12.7    & R12  \\
L32 & $\rm MM\,J104539.6+585419$ & 161.41579 & 58.906917 & 2.40        & P-IR  & M10 & 12.57  & M10   \\
L33 & $\rm MM\,J104535.5+585044$ & 161.39608 & 58.847139 & 2.63        & P-IR  & R12 & 12.8     & L12  \\
L34 & $\rm MM\,J104453.7+585838$ & 161.22346 & 58.978194 & 2.46        & P-IR  & R12 & 12.6    & L12   \\
L35 & $\rm MM\,J104717.9+585523$ & 161.82546 & 58.923833 & 2.14        & P-IR  & R12 & 12.9     & L12  \\
L36 & $\rm MM\,J104608.1+590744$ & 161.53421 & 59.129444 & 4.5          & CY     & L11 & 12.5     & L12   \\
L37 & $\rm MM\,J104610.8+585242$ & 161.54575 & 58.879139 & 1.72        & P-IR   & M10 & 12.81 & M10     \\
L38 & $\rm MM\,J104444.5+590817$ & 161.18704 & 59.138361 & 3.6          & CY     & L11 & 12.5      & L12   \\
L39 & $\rm MM\,J104611.9+590231$ & 161.55012 & 59.042583 & 2.59        & P-IR  & R12 & 12.5      & L12  \\
L40 & $\rm MM\,J104658.7+590633$ & 161.74304 & 59.109306 & 0.78        & P-IR  & R12 & 12.5      & L12  \\
L41 & $\rm MM\,J104600.7+585502$ & 161.50129 & 58.917944 & 1.49        & P-IR  & R12 & 12.6    & L12
\enddata
\label{t-targets}
\tablenotetext{a}{Position of $20\,\rm cm$ radio counterpart}
\tablenotetext{b}{Redshift type: S = spectroscopic, P-IR = $Herschel$-based photometric, P-O = optical-based
photometric, CY = estimated using the spectral index $\alpha^{20\,\rm cm}_{850\,\rm\mu m}$ \citep{cari99}.}
\tablenotetext{c}{References: P06 = \citet{poll06}, O09 = \citet{owen09}, F10 = \citet{fiol10}, M10 = \citet{magd10}, S10 = \citet{stra10}, L11 = \citet{lind11}, R12 = \citet{rose12}, L12 = This Work}
\tablenotetext{d}{8--$1000\,\rm\mu m$}

\end{deluxetable}

\begin{deluxetable}{cccc}
\tabletypesize{\scriptsize}
\tablecaption{redshift uncertainties}
\tablewidth{0pt}
\tablehead{\colhead{Redshift type} & \colhead{N galaxies} & \colhead{$\Delta z$} & 
\colhead{References} }
\startdata
\multirow{3}{*}{Spectroscopic} & \multirow{3}{*}{3} & \multirow{3}{*}{$\simeq 0$} & \citet{poll06} \\
&  &  & \citet{fiol10} \\
&  &  & \citet{owen09} \\ \\
Optical-based photometric & 3 & 0.2 & \citet{stra10} \\ \\
\multirow{2}{*}{Infrared-based photometric} & \multirow{2}{*}{23} & \multirow{2}{*}{0.4} & \citet{magd10} \\
 & & & \citet{rose12} \\ \\
Spectral index-based estimate & \multirow{2}{*}{9} & \multirow{2}{*}{0.6} & \multirow{2}{*}{\citet{lind11}} \\
\citep{cari99} & & &
\enddata
\label{t-zerrors}
\end{deluxetable}

\begin{deluxetable}{cccccc}
\tabletypesize{\scriptsize}
\tablecaption{X-ray Stacking Results}
\tablewidth{0pt}
\tablehead{
\colhead{Band}                     & 
\colhead{Energy}                     & 
\colhead{Net Rate}                 & 
\colhead{Flux ($\Gamma=1.6$)}      & 
\colhead{Flux ($\Gamma=1.9$)}      & 
\colhead{$L_{\rm X} \,(\Gamma=1.6)$}   \\ 
& keV &
$10^{-6}\,\rm s^{-1}$                 & 
$10^{-17}\,\rm erg\,cm^{-2}\, s^{-1}$  & 
$10^{-17}\,\rm erg\,cm^{-2}\, s^{-1}$  & 
$\rm 10^{42}\,erg\,s^{-1}$}

\startdata
$S_{C}$ &  0.5--2.0                & $  8.0^{+ 2.1}_{- 2.0_{ }}  $  & $4.7^{+1.2}_{-1.2}$ & $4.9^{+1.3}_{-1.2}$  & -- \\
$H_{C}$  &      2.0--8.0           & $  1.2^{+3.2}_{-2.9_{}}    $  & $<9.7$ & $<9.1$ &-- \\
$B_{C}$   &          0.5--8.0       & $  9.2^{+ 3.8}_{- 3.5_{ }}  $  & $10.1^{+4.2}_{-3.9}$ & $9.0^{+3.7}_{-3.4}$   & -- \\
$H_{C}^{\rm rest}$& 0.55-2.22  & $  8.0^{+ 2.1}_{- 2.3_{}}  $  & $4.8^{+1.3}_{-1.4}$ & $4.9^{+1.3}_{-1.4}$ &  $3.0\pm 1.1$\\
$H^{\rm rest}$ & 0.55-2.78   & $  8.6^{+ 2.4}_{- 2.5}  $  & $5.7^{+1.6}_{-1.7}$ & $5.7^{+1.6}_{-1.7}$ & $ 3.6\pm 1.3$  
\enddata

\label{t-data}
\tablecomments{Unabsorbed fluxes calculated assuming the given photon index
with Galactic absorption only.  Hydrogen column density taken as that of the 
central {\it Chandra} pointing, $N_{H}=6.6\times 10^{19}\,\rm cm^{-2}$ \citep{star92}.
Luminosity calculation uses $\Gamma=1.6$.}
\end{deluxetable}

\end{document}